\documentclass[reprint, superscriptaddress, amsmath,amssymb, aps, pra, longbibliography]{revtex4-1}

\usepackage{siunitx}

\usepackage{soul}
\usepackage{amsmath} 
\usepackage{amssymb}
\usepackage{mathtools}
\usepackage{graphicx}
\usepackage{dcolumn}
\usepackage{bm}
\usepackage{amsmath}
\usepackage{graphicx}
\usepackage{amsfonts}
\usepackage{subfigure}
\usepackage{graphicx}
\usepackage{array}
\usepackage{float}
\usepackage{color}
\usepackage{multirow}
\usepackage{amssymb}%
\usepackage[colorlinks=true,linkcolor=blue]{hyperref}%
\hypersetup{allcolors=blue}
\usepackage[normalem]{ulem}
\usepackage{xcolor}

\newcommand{\bra}[1]{\ensuremath{\left\langle #1 \right\vert}}
\newcommand{\ket}[1]{\ensuremath{\left\vert #1 \right\rangle}}
\usepackage{mathtools}
\DeclarePairedDelimiterX\braket[2]{\langle}{\rangle}{#1 \delimsize\vert #2}

\begin{document}

\title{Sagnac Tractor Atom Interferometer on Photonic Integrated Circuit}

\date{\today }

\author{Lefeng Zhou}
 \email{lefengz@umich.edu}
 \affiliation{Department of Physics, University of Michigan, Ann Arbor, MI 48109, USA}

\author{Anne Graf}
 \affiliation{Department of Physics, University of Michigan, Ann Arbor, MI 48109, USA}

\author{Georg Raithel}
 \affiliation{Department of Physics, University of Michigan, Ann Arbor, MI 48109, USA}

\begin{abstract}

We study the theory of, and propose an experimental design for, a Sagnac tractor atom interferometer based on a photonic integrated circuit (PIC). The atoms are trapped in counter-rotating azimuthal optical lattices, formed by interfering evanescent fields of laser modes injected into circular PIC waveguides. We develop quantum models for the radial and azimuthal dynamics of the interfering atoms in adiabatic frames, which provide computational efficiency. The theory is applied to an exemplary PIC, for which we first compute field modes and atom trapping potentials for $^{87}$Rb. We then evaluate non-adiabaticity, fidelity, and sensitivity of the exemplary PIC.

\end{abstract}

\maketitle

\section{Introduction}
\label{sec:intro}

In atom interferometry (AI) \cite{Cronin2009-cp}, matter-wave interference of cold atoms is used to test fundamental principles of physics~\cite{PhysRevLett.112.203002, Superpos, Jaffe2017-oe, Rosi2017-tj} and to perform precise measurements of fundamental constants~\cite{Fixler2007-aq, Hanneke2008-fz, Morel2020-vy}. Using massive atoms instead of photons leads to short de Broglie wavelengths, as compared to optical wavelengths. This also leads to high sensitivity to gravity and inertial forces; as a result, AI sensors are also suitable for gravimetry~\cite{menoret2018gravity} and inertial sensing~\cite{garrido2019compact}.

AI can be applied to measure both linear acceleration~\cite{Jaffe2017-oe, Rosi2017-tj, Fixler2007-aq} and rotation~\cite{Dickerson2013-pa, Moan2020-qm, gautier2022accurate, Wu2007-lj}. Rotation sensing relies on the Sagnac phase shift that accumulates due to the rotation of the interferometer platform or instrument against an inertial frame. The rotation sensitivity scales with the area enclosed by the interferometer paths. A multitude of schemes have been proposed and/or realized to increase the enclosed area, such as guiding atoms around a closed loop multiple times~\cite{Beydler2024-jz, Wu2007-lj, Campbell2017-kk}.

The recently proposed tractor atom interferometry (TAI) \cite{Raithel_2023, Duspayev2021-xi} is a good candidate to achieve high sensitivity with a small footprint. In TAI, atoms are trapped in three-dimensional (3D) tractor potentials throughout the entire AI sequence. The atoms are shuttled in the 3D-confining tractor traps along user-programmed paths, which define the AI configuration. As such, in TAI the atoms have zero external spatial degrees of freedom. This contrasts with free-space \cite{Asenbaum2017-ef, Rosi2014-yu, Xu2019-cn}, point-source \cite{Dickerson2013-pa, Hoth2016-yk, Chen:20}, and guided-wave AI \cite{Wu2007-lj, Moan2020-qm}, where the atoms are free to move along one spatial coordinate, at a minimum. The 3D trapping in TAI ensures closure, and it enables long holding times and complex multi-loop paths. In previous work~\cite{RotationSensing}, we have considered Sagnac TAI in counter-rotating azimuthal optical lattices, formed by interfering multi-frequency Laguerre-Gaussian modes.

In many applications, the size, weight, and power consumption (SWaP) exert limits on the interferometer design. Because atom-nanophotonic integration could reduce SWaP, there have been inroads in that area. Atoms have been successfully loaded and trapped in the vicinity of photonic integrated circuits (PICs) for studying quantum electrodynamics and quantum information science with optical tweezers \cite{menon2024integrated, thompson2013coupling}, optical conveyors \cite{xu2023transporting, kim2019trapping}, and optical funnels \cite{zhou2023coupling}. Atom guides on PIC platforms, which emulate the light-guiding function of optical fibers to guide atomic waves, were studied in~\cite{barnett2000substrate, ovchinnikov2020towards}.

In this paper, we propose a Sagnac TAI on a compact PIC platform, with the goal of creating a rotation-sensing device that offers a favorable combination of low SWaP and high rotation sensitivity. We design an optical lattice on a PIC ring resonator, which shuttles the atoms in 3D traps hovering above the ring. We first formulate a theory describing the radial and azimuthal dynamics of atoms trapped in a circular optical lattice. We reduce analytical and numerical complexity using several approximations, which we will justify. Then, we design a multi-mode PIC waveguide suited for the implementation of counter-rotating azimuthal optical lattices using a combination of attractive and repulsive optical potentials. We then proceed with quantum simulations that show a sensitivity of several nrad/s, which can be achieved with a PIC ring radius of $R_0 = \SI{600}{\um}$, many-loop tractor-trajectory operation, and interrogation times of the instrument rotation of around $\SI{1}{s}$.

\section{Sagnac TAI on a Photonic Chip - Basic concepts}
\label{sec:concepts}

\begin{figure}[htb]
 \centering
 \includegraphics[width=0.45\textwidth]{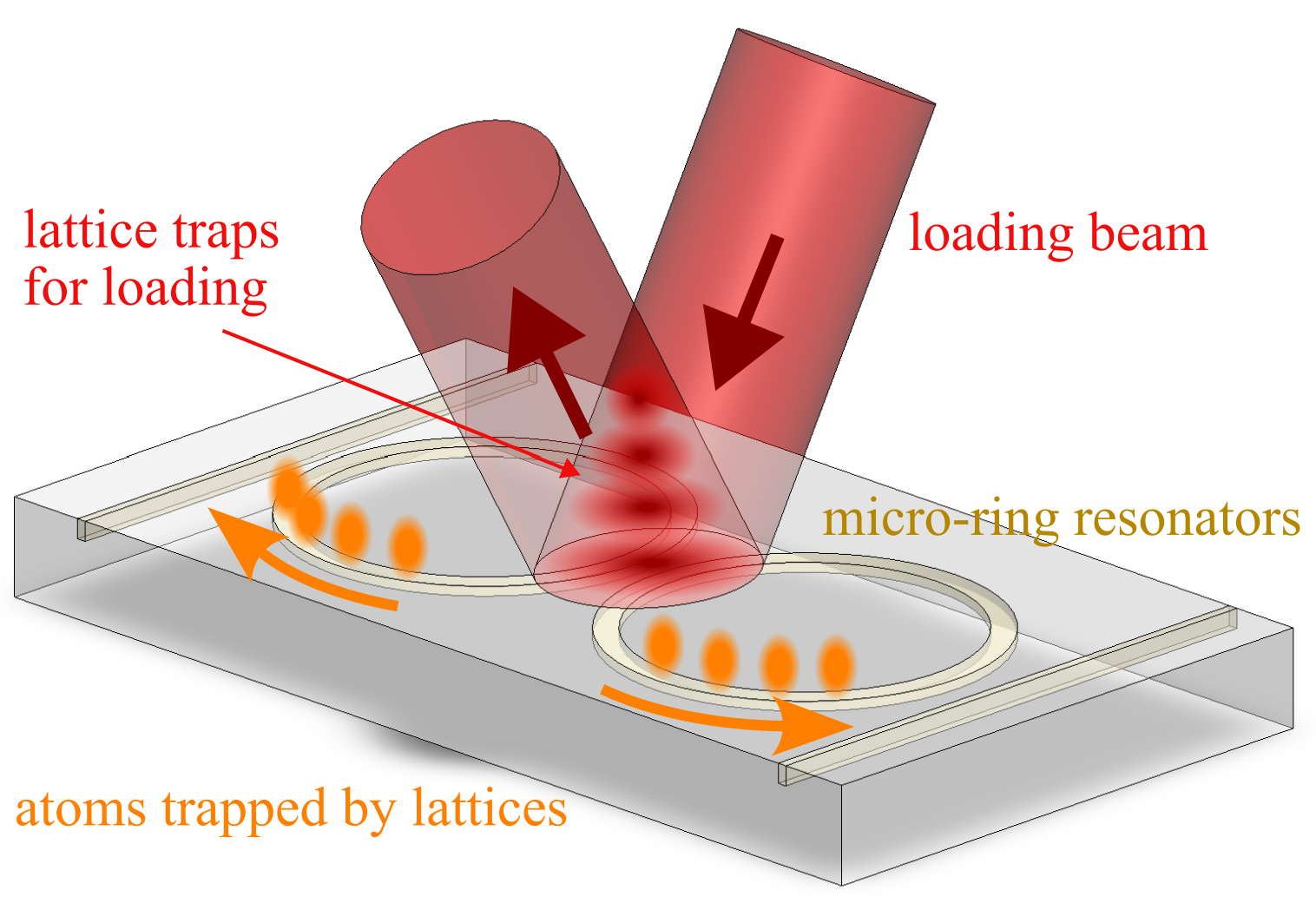}
 \caption{Two micro-ring resonators side by side on a photonic chip. The loading beam is partially reflected by the top surface of the chip, forming pancake-shaped traps that coherently load the azimuthal lattices above the ring resonators with cold atoms. Details are described in the text.}
 \label{fig:chip_and_loading_beam}
\end{figure}

In TAI, the atomic wave-packet components are split, shuttled along predetermined paths, and recombined, while being confined in 3D tractor potentials at all times. The tractor potentials are programmed such that closure is guaranteed and the split paths enclose a non-zero area. Here we consider rotating azimuthal 1D optical lattices that shuttle $^{87}$Rb atoms on circular tractor trajectories with opposite directions, and that are red-detuned relative to the Rb D lines (780~nm and 795~nm). Azimuthal lattices can be created using micro-ring resonators on a photonic chip, as  visualized in Fig.~\ref{fig:chip_and_loading_beam}. Each micro-ring resonator is bi-directionally pumped through a linear waveguide section, forming an azimuthal standing wave along the ring. Changing the relative phase of the counter-propagating resonator modes results in azimuthal translation of the lattice anti-nodes along the micro-ring resonator. A fixed frequency difference between the modes causes an azimuthal rotation at a fixed angular velocity. The Sagnac TAIs in Figs.~\ref{fig:chip_and_loading_beam} and~\ref{fig:ring} utilize two synchronized counter-rotating azimuthal lattices in two PIC ring resonators (or PIC-TAILs for Photonic-Chip Tractor Atom Interferometer optical Lattices). The interference signal, produced upon recombination of wave-packet components traveling in corresponding pairs of wells in the counter-rotating lattices, contains information on the angular velocity of the entire TAI apparatus against an inertial frame.

In Fig.~\ref{fig:chip_and_loading_beam}, the counter-rotating PIC-TAI loops have equal areas that extend on a planar chip surface. The loops are traversed by split wave-packet components, which are localized in azimuthal PIC-TAIL sites that move in clockwise (cw) and counter-clockwise (ccw) directions at identical rotation speeds. According to the semiclassical Feynman path integral formalism~\cite{Storey1994} the AI phase difference, $\Delta\phi_S$, accumulated by the split wave-packet components, is given by
\begin{align}
 \Delta\phi_S &= \frac{1}{\hbar}\int_0^{T} dt\ (L^+ - L^-) \nonumber \\
 &= \frac{2 M K \bold A \cdot \bold\Omega}{\hbar}
\quad . 
\label{eq:semiclassical}
\end{align}
There, $L^\pm$ are the Lagrangians of the two interferometric paths. Under symmetric operation of the PIC rings, the potential energy is the same for the two paths, and the difference of the Lagrangians is only due to the difference in kinetic energy induced by the instrument rotation. The PIC-TAI rings have an area of $|{\bf{A}}|$ each, and their area vectors are pointing in opposite directions. The instrument's angular velocity against an inertial frame, ${\bf{\Omega}}$, is averaged over the time the wave-packets are split between the PIC-TAI loops. The mass of one atom is denoted $M$. The counting index $K$ denotes the number of half rotations in the rings ({\sl{i.e.}} one full rotation in each loop corresponds to $K=2$). Closure of the TAI scheme in Fig.~\ref{fig:chip_and_loading_beam} requires an even, integer value of $K$. We will show in Sec.~\ref{sec:sims}~B that the semiclassical TAI phase in Eq.~(\ref{eq:semiclassical}) agrees with a fully quantum-mechanical simulation, given sufficient adiabaticity.

For ${\bf{A}} \cdot {\bf{\Omega}} = A \, \Omega$, a change in the AI phase, $\Delta\phi_S$, by $2\pi$ corresponds to a change in the instrument's rotation rate by
\begin{align}\label{eq: feynman}
 \Omega_0 = \frac{\pi\hbar}{K M A} \quad.
\end{align}
Denoting the uncertainty of the $\Delta\phi_S$-measurement by $\delta \phi_S$, the signal-to-noise ratio ($\mathrm{SNR}$) of the AI phase is $2 \pi / \delta \phi_S$, and the precision of rotation measurement then equals 
\begin{equation}
\delta \Omega = \Omega_0 / \mathrm{SNR} = \frac{\delta \phi \,\, \hbar}{2 \, K \, M \,A} \quad .    
\label{eq:sensitivity}
\end{equation} 
In the proposed PIC-TAI implementation, the area $A$ will be limited to $\lesssim 1$~mm$^3$. High sensitivity must therefore be derived from high azimuthal-lattice rotation speeds, which allow us to obtain $K$-values of $\sim 1000$ over the targeted wave-packet splitting times of $\sim \SI{1}{s}$. Without squeezing~\cite{spinsqueezing}, the $\mathrm{SNR}$ in Eq.~(\ref{eq:sensitivity}) is limited by quantum projection noise (QPN), for which $\delta \phi_S \sim 1/ \sqrt{N}$ and the $\mathrm{SNR} \sim 2 \pi \sqrt{N}$, where $N$ is the total number of interrogated atoms. Here, we may anticipate using $\sim 1000$ lattice sites in parallel, occupied with coherent fragments of a Bose-Einstein condensate (BEC) of $\sim 10$ atoms each, leading to a $\delta \phi \sim 0.01$ and a $\mathrm{SNR} \sim 2 \pi \times  100$ in Eq.~(\ref{eq:sensitivity}).

The PIC-TAI approach must be embedded within the strategy for loading, splitting, and recombination of the atomic wave-packets between the pairs of counter-rotating PIC-TAILs. These functions can be provided via an auxiliary loading beam that is focused on the top of the photonic chip, at an angle of incidence $\lesssim 20^\circ$ against the normal, into the loading area, where the two rings are closest to each other. About 5$\%$ of the beam power
is reflected, forming a loading optical lattice with a modulation depth along the chip normal of about 50$\%$ of the maximum lattice depth (see Fig.~\ref{fig:chip_and_loading_beam}). The auxiliary lattice consists of pancake-shaped traps with a Gaussian beam parameter of the loading beam of $w_0 \sim \SI{5}{\um}$, extending into the plane parallel to the chip surface. The traps are stable against frequency and phase fluctuations of the loading beam due to proximity to the chip surface. The pancake trap closest to the chip overlaps with both counter-rotating PIC-TAILs. Depth and contrast of the pancake traps increase with increasing proximity to the chip, which is conducive to preferential build-up of BEC population in the pancake trap closest to the chip. The fraction of the BEC prepared in the closest and deepest pancake trap is coherently coupled into pairs of azimuthal PIC-TAIL sites via adiabatic transfer from the pancake trap into the adiabatically turned-on PIC-TAILs. During the loading, the PIC-TAIL wells are at rest in the instrument frame and are azimuthally aligned for symmetric loading of phase-coherent wave-packet components into pairs of wells. This loading process thus performs the same function as beam splitters do in other types of interferometers.
After coherent loading from the BEC, the PIC-TAILs are adiabatically spun up in a counter-rotating fashion and are kept spinning at a fixed rotation rate for some period of time to accumulate a Sagnac AI signal. Subsequently, the PIC-TAILs are decelerated back down to rest in the instrument frame. Recombination occurs by adiabatic transfer of the atoms back into the loading lattice. The spatial structure of the interference pattern formed by the recombined wave-packets carries the information of the AI phase $\Delta \phi_S$.

Important aspects of the thusly described loading approach are as follows: the loading occurs from a single, coherent BEC source, and the azimuthal angles of the counter-rotating PIC-TAILs are exactly matched. Azimuthal phase matching can be accomplished by deriving the optical-lattice field modes from a common laser using RF components and electro-optics with phase noise of $\sim \SI{1}{mrad}$. These methods will be conducive to a high degree of symmetry of the PIC-TAILs in terms of trap depth and azimuthal phase, translating into low noise in differential trap depth and low azimuthal phase mismatch between the lattices. The described loading protocol may be refined further without changing the overall concept.

Loading atoms into photonic devices, though being critical and experimentally challenging, does not affect the atomic dynamics in the azimuthal PIC-TAILs or during the sensing stage, which is the main focus of the present work. The exemplary loading scheme, laid out above, is described for a scalar TAI implementation, in which the wave-packet components are coherently split between spatially separated tractor wells trapping the same internal atomic spin state. The AI phase information is contained in the vibrational state of the wave-function after recombination~\cite{Duspayev2021-xi}.

Our numerical modeling of the Sagnac phase accumulation,  presented in Sec.~\ref{sec:sims}, also covers the case of spinor-TAI~\cite{Duspayev2021-xi}, where the atomic wave-function is split between different internal spin states. In spinor-TAI, different spin states are confined in spin-dependent tractor traps, which may or may not overlap in space. The atomic dynamics in the PIC-TAILs during the Sagnac phase accumulation stage, which are the focus of the present paper, are the same for both scalar and spinor interferometers. Detailed simulations of PIC-TAI loading and recombination sequences for specific cases may be the subject of future work.

\section{Quantum Dynamics in a Rotating Azimuthal Optical Lattice}
\label{sec:dynamics}

\subsection{Rotating Reference Frames}

Before proceeding with our proposed realization of a Sagnac PIC-TAI, we develop the theoretical methods to be used to estimate the performance  
characteristics of such a device. In the following, there are three relevant cylindrical frames: an inertial frame, the instrument frame, and frames that are co-rotating with the azimuthal PIC-TAILs (``lattice frames''). The rotation axes of the instrument and lattice frames are assumed to coincide with, and point along, $\hat{\bf{z}}$, the $z$-axis of the inertial frame. The lattices then extend in the cartesian $xy$-plane in any of the frames. The cylindrical coordinates $(r, \theta, z)$ with $r=\sqrt{x^2+y^2}$ are in the lattice frame. The angular velocity of the rotating azimuthal PIC-TAILs against the instrument frame is denoted ${\boldsymbol{\omega}} (t) = \pm \hat{\bf{z}} \omega(t)$, with a positive ramp function $\omega(t)$. The instrument's angular velocity against the inertial frame is denoted ${\boldsymbol{\Omega}} = \hat{\bf{z}} \Omega (t)$. The instrument rotation is very slow relative to the peak value of the ramp function $\omega(t)$, and therefore considered to be quasi-static. The angular velocities of the rotating PIC-TAILs against the inertial frame are ${\boldsymbol{\omega}}_L = \hat{\bf{z}} (\pm \omega(t) + \Omega)$. 

The axial dynamics are frozen, {\sl{i.e.}}, it is assumed that the wave-packets are always in the ground state of the motion along $z$. This is justified by the fact that the trapping potentials transverse to the chip surface are tighter than those in the radial direction (see Figs.~\ref{fig:potential_cross_section}-\ref{fig:potential_xy}). Also, for the present work one may assume that there is no significant platform acceleration. If axial dynamics were included, they would couple to the radial and azimuthal degrees of freedom via the optical trapping potential, $V(r,\theta,z)$, which is, typically, not exactly separable. Notably, none of the fictitious forces discussed in the following would couple to the axial motion. 

In the lattice frames, {\sl{i.e.}}, in frames that co-rotate with an azimuthal PIC-TAIL, the atoms are subject to three fictitious forces. The centrifugal force, Coriolis force, and Euler force have the following forms in the plane perpendicular to our angular velocity ${\boldsymbol{\omega}}_L$:

\begin{align}
 \mathbf{F}_r &= M\omega_L^2 \, r \, \bold{\hat r} \nonumber \\
 \mathbf{F}_c &= 2M\omega_L \, (r\dot\theta \bold{\hat r} - \dot r \boldsymbol{\hat\theta}) \nonumber \\
 \mathbf{F}_e &= -M\dot\omega_L r \boldsymbol{\hat\theta}
 \label{eq:ff}
\end{align}

The relevant atoms that give rise to the Sagnac AI signal co-rotate with the lattice and are trapped in states close to the instantaneous ground states of the rotating azimuthal lattice potential. In the adiabatic limit, the majority of atoms reside in such states. Hence, in the adiabatic basis, which is spanned by the stationary states of the co-rotating lattice frame, the relevant atomic states are superpositions of only a small number of states, and the Hilbert space of relevant base kets is quite small. As such, solving the azimuthal quantum dynamics in the co-rotating lattice frame is most efficient. 

The radial and azimuthal dynamics are affected by the fictitious forces, which require some attention. The Coriolis force couples radial and azimuthal dynamics. However, the Coriolis coupling will be seen to be relatively weak compared to other forces in Secs.~\ref{sec:dynamics}~B and C, and it is neglected in the present work. Moreover, the optical trapping potential near the trap minima almost separates in the radial, azimuthal and axial degrees of freedom, and the dynamics in different degrees of freedom have considerably different time scales. The axial dynamics are already assumed to be frozen. Due to the sum of these features, radial and azimuthal dynamics adiabatically separate and can be treated independently, as done in the following.

The Hamiltonian in the lattice frame is
\begin{align}\label{eq: Hrf}
 H = \frac{p_r^2}{2M} + \frac{L_z^2}{2Mr^2} + V(r,\theta) - \omega_L (t) L_z
\end{align}
where $r$ and $p_r$ denote radial position and momentum, and $\theta$ and $L_z$ denote azimuthal angle and angular momentum. The angular velocities of the ccw and cw-rotating lattices relative to the inertial frame are $\omega_L(t) = \pm \omega(t) +\Omega$, with the above definitions of the angular velocities of the frames. Over the course of the Sagnac-TAI sequence, the ramp function $\omega(t)$ is ramped up from $0$ to a peak value, $\omega_s$, which is on the order of 1~kHz, then held steady for a time on the order of 1~s, and finally ramped back down to 0 for closure of the TAI. The instrument's angular velocity relative to the inertial frame, $\Omega \ll \omega_s$, is considered fixed during the Sagnac-TAI sequence.

In the approximation that radial and azimuthal dynamics adiabatically separate, the trapping potential is $V(r,\theta) \approx V_r(r) + V_\theta(\theta)$: the atoms are tightly confined in the vicinity of the PIC ring resonator's radius $R_0$, and radial and azimuthal dynamics are not coupled. The Hamiltonians in the radial and azimuthal directions can then be written as
\begin{align}\label{eq: H0}
 H_r &= \frac{p_r^2}{2M} + V_r(r)
 \end{align}
 \begin{align}\label{eq: HB}
 H_\theta &= \frac{L_z^2}{2I} + V_\theta(\theta) - \omega_L L_z
\end{align}
where $I = M R^2$ is the moment of inertia of one atom. There, the radial equilibrium position $R$ is mostly given by the radius of the PIC resonators, with a mild increase caused by the centrifugal force, as discussed in the next section. In the following, we derive equations from Eqs.~(\ref{eq: H0}) and ~(\ref{eq: HB}) that will be used in Sec.~\ref{sec:sims} to simulate the quantum dynamics in a PIC-TAI sample case.

\subsection{Radial Dynamics}

We assume that the radial light-shift trapping potential for the atoms is harmonic near its bottom, $V_r(r) = \frac{1}{2} M \omega_r^2 (r-R_0)^2$. There, $\omega_r$ is the radial trap frequency and $R_0$ is the radial position of the trap minimum under the absence of azimuthal rotation, which practically coincides with the PIC resonator's radius. 
Over the range of conditions of interest, the radial quantum state remains near the quantum ground state of the radial motion, which is a requirement to produce high-visibility TAI signals. As such, the harmonic approximation is valid  under the conditions of interest.

The radial state is perturbed by the time-dependent centrifugal force, $F_r = M \omega_L^2 r$, which is accounted for by  adding a radial potential $ - M \omega_L^2 r^2 /2$ in Eq.~(\ref{eq: H0}). The centrifugal force $F_r = M\omega_L^2 r$ slightly increases the radius of the interferometric path relative to $R_0$ by an amount denoted $\delta R(t)$. In the lowest order, the centrifugal translation of the radial potential minimum is
\begin{align}\label{eq: x0}
 \delta R(t) = R(t) \frac{\omega_L^2(t)}{\omega_r^2} \quad. 
\end{align}
Here, we note that relatively tight radial trapping, $\omega_r \gg \omega_L$, is important to keep the atoms on a designated tractor path with minimal radial sag $\delta R(t)$, {\sl{i.e.}} $\delta R \ll R_0$. This is achieved through a large force constant of the radial trap, $M \omega_r^2$.
In the harmonic approximation, and defining $x = r - R_0$, the radial Hamiltonian is
\begin{align}\label{eq: Hr}
 \hat{H}_r = \frac{\hat{p}_r^2}{2M} + \frac{1}{2}M\omega_r^2 \hat{x}^2 - M \omega_L(t)^2 R(t) \hat{x} \quad
\end{align}
where we have added hats that indicate which variables are operators in the subsequent quantum analysis. 
Since the radial sag $\delta R \ll R_0$ in all cases of interest, in the centrifugal-potential operator [the last term in Eq.~(\ref{eq: Hr})] we may assume a fixed $R(t) = R_0$ and write $M \omega_L(t)^2 R_0 \hat{x} $. The time dependence of $\hat{H}_r$ is then given by the angular velocity of the azimuthal lattice-rotation, $\omega_L(t) = \pm \omega(t) + \Omega$. The ramp function $\omega(t)$ ranges up to several krad/s and outweighs the platform rotation rate $\Omega$ by multiple orders of magnitude. For the establishment of radial adiabaticity we may therefore also set $\omega_L(t) = \omega(t)$. Under these approximations,
\begin{align}\label{eq: Hr2}
 \hat{H}_r - E_{r0}(t)= \frac{\hat{p}_r^2}{2M} + \frac{1}{2}M\omega_r^2 (\hat{x} - \delta R(t))^2 \quad
\end{align}
with an inconsequential global time-dependent energy shift $E_{r0} = - R_0^2 M \omega_L^4(t)/\omega_r^2$.

In the rotating frame, atoms trapped in the rotating azimuthal lattice potential move at angular velocities $|\dot\theta| \ll |\omega_L|$, {\sl{i.e.}}, the azimuthal atomic velocity in the rotating frame is much slower than the azimuthal velocity of the lattices in the lab and inertial frames. From Eq.~(\ref{eq:ff}), it follows that the Coriolis force in the radial direction is much smaller than the centrifugal force, which serves as one of our justifications for discarding the Coriolis effects for now. 

A particle in a harmonic trap is quantized with ladder operators. If the trap is fixed in position, the quantum problem is solved using a basis of fixed (diabatic) number states, $\{\ket{n}, n = 0,1,2,\cdots\}$. However, working in the diabatic basis becomes inefficient if the radial shift $\delta R(t)$ in Eq.~(\ref{eq: Hr2}) becomes much larger than the standard position uncertainty $\sqrt{\hbar / (M \omega_r)}$. In that case, the number of required basis states would exceed $R_0^2 M \omega_L^4(t)/(\hbar \omega_r^3)$.

Therefore, we work in the adiabatic basis, defined by  
\begin{align}\label{eq: translatestate}
 \ket{n(t)} = \hat{T}(\delta{R}(t))\ket{n}
\end{align}
with the translation operator $\hat{T} (\delta R(t)) = \exp (- i\hat{p}_r \delta R(t)/\hbar)$. The translated states, $\{\ket{n(t)}, n = 0,1,2,\cdots\}$, then are the instantaneous eigenstates of $\hat{H}_r$, {\sl{i.e.}} they form the adiabatic basis (with $\omega_r$ assumed to be fixed). The adiabatic energies are $E_n = (n+1/2) \hbar \omega_r$, up to the irrelevant global shift $E_{r0}(t)$ in Eq.~(\ref{eq: Hr2}). Denoting 
\begin{equation}
    \vert \psi(t) \rangle = \sum_{n=0}^\infty c_n(t) \vert n(t) \rangle
    \quad ,
\end{equation}
the state's coefficients in the adiabatic basis follow the equations
\begin{align}\label{eq: rseq}
 \dot c_n = \frac{E_n}{i\hbar} c_n - \sum_{n'} c_{n'} \bra{n(t)} \partial_t \ket{n'(t)} \quad.
\end{align}
The non-adiabatic couplings on the RHS of Eq.~(\ref{eq: rseq}) are evaluated at time $t$ and are given by 
\begin{align}\label{eq: derivativesimplify}
 \bra{n(t)} \partial_t\ket{n'(t)} &= \bra{n} \hat{T}^\dagger (t) \partial_t \left[ \hat{T} (t) \ket{n'} \right]
 \nonumber \\
~&= \bra{n} \hat{T}^\dagger (t) \left[\partial_t \hat{T} (t) \right] \ket{n'}
 \nonumber \\
~&= \bra{n} \hat{T}^\dagger (t)\left[ \frac{d}{dt} \delta R(t) \right] \frac{ - i \hat{p}_r}{\hbar} \hat{T} (t) \ket{n'}
 \nonumber \\
~&= \frac{ - i}{\hbar} \left[ \frac{d}{dt} \delta R(t) \right]  \bra{n} \hat{p} \ket{n'} \quad,
\end{align}
where we write $\hat{T}(t)$ for $\hat{T}(\delta R(t))$ for brevity. Eq.~(\ref{eq: rseq}) then becomes
\begin{align}\label{eq: rseqsimplify}
 \dot c_n = \frac{E_n}{i\hbar} c_n - \sqrt{\frac{M\omega_r}{2\hbar}} \left[ \frac{d}{dt} \delta R(t) \right]  \left(\sqrt{n}c_{n-1} - \sqrt{n+1}c_{n+1}\right)
\end{align}
with
\begin{align}\label{eq: x0dot}
\frac{d}{dt} \delta R(t) = 2 R_0 \frac{\omega_L \dot \omega_L}{\omega_r^2} \quad .
\end{align}
Only neighboring adiabatic states of the radial motion become coupled. Since the non-adiabatic coupling is proportional to the angular acceleration $\dot\omega_L$, non-adiabatic effects arise only during the time intervals when the azimuthal lattices are spun up and back down. In cases of interest, the atoms are initialized in the ground state and non-adiabatic effects are weak, so that typically less than 10 adiabatic states are sufficient. As such, performing the computation in the adiabatic basis reduces the required computation power. Moreover, the $|c_n|^2 (t)$ obtained by solving Eq.~(\ref{eq: rseqsimplify}) are equivalent to the number-state populations in the frame of reference of the shifting radial potential, allowing a direct evaluation of the severity of non-adiabatic excitation.

\subsection{Azimuthal Dynamics}

The Hamiltonian of a particle in a circular optical lattice in the co-rotating frame~\cite{RotationSensing} is:
\begin{align}\label{eq: Htheta}
 \hat{H}_\theta &= \frac{\hat{L}_z^2}{2I} + V_0\cos(m \hat{\theta}) - \omega \hat{L}_z
\end{align}
where $I=M R_0^2$ is the moment of inertia, $V_0$ half the lattice depth, and $m$ the number of lattice sites along the azimuthal PIC-TAILs of interest. For simplicity, we assume $m$ to be even. The effect of the Coriolis force in the azimuthal direction, shown in Eq.~(\ref{eq:ff}), is proportional to $\dot r$, which scales with $d \delta R(t) / dt$. Since $d \delta R(t) / dt$ is relatively small due to the tight radial confinement, the Coriolis coupling is neglected in the recent work.

We can choose the lattice-frame Bloch states to span the Hilbert space,
\begin{align}\label{eq: blochstate}
 \psi_l^n(\theta) & = \langle \theta \vert \psi_l^n \rangle= \exp(i l\theta) u_l^n(\theta) \nonumber \\
 & =
 \sum_{k} c_{l,k}^n\exp(i(km + l)\theta)
\end{align}
where $l$ is an integer index ranging from $-m/2$ to $m/2-1$, which is equivalent to the quasi-momentum, $q = 2 (l/m) (\pi/a_L) = l/R_0$, with lattice period $a_L = 2 \pi R_0/ m$.
The Hamiltonian in Eq.~(\ref{eq: Htheta}) does not couple states of different $l$. Also, the quasi-momentum on a circle with $m$ wells is a discrete quantum number with $m$ values. The Bloch functions $u_l^n(\theta)$ have a periodicity of $2\pi/m$, the same as $H_\theta$. The integer quantum number $n=0,1,2,...$ is a band index that labels the eigen-energies $E_l^n$ in ascending order. The integer quantum number $k$ denotes the free-particle states $\vert k \rangle$ in the co-rotating lattice frame, which have a periodicity of $2 \pi$ and are normalized as $\langle \theta \vert k\rangle = \exp(i k \theta)/\sqrt{2 \pi}$. The Bloch-state coefficients in Eq.~(\ref{eq: blochstate}), $c_{l,k}^n$, are obtained by expressing the Hamiltonian Eq.~(\ref{eq: Htheta}) in the $\vert k \rangle$-basis and solving 
\begin{multline}\label{eq: aes}
 \left[\frac{\hbar^2}{2I}(km + l)^2 - \hbar\omega_L (km + l) - E_l^n \right] c_{l,k}^n \\
 + \frac{1}{2}V_0(c_{l,k+1}^n + c_{l,k-1}^n) = 0
\end{multline}
Solving Eq.~(\ref{eq: aes}) yields all eigen-energies $E_l^n$ and -states for the integer quasi-momentum equivalent $l$, with integer band index $n$. The (discrete) band structure is obtained by separately solving Eq.~(\ref{eq: aes}) for all integers from $l=-m/2$ to $l=m/2 -1$. 

To numerically solve Eq.~(\ref{eq: aes}), we truncate the $k$-range to $ |k| \le k_\text{max}$. This is appropriate because $k$-states with very large $k$ have too high of an energy to significantly contribute to the Bloch states that are bound in the lattice. Minimizing the diagonal term in Eq.~(\ref{eq: aes}) with respect to $k$ and noting that $l$ contributes only a relatively small amount of energy for large $m$ (our case), we see that the range of essential $k$-states is centered around $k_0 =  I\omega_L/(\hbar m)$. We may further estimate the half-range of $k$-states that contribute to the lattice-bound states, $\Delta k$, by $(\hbar m k)^2/(2I) \approx V_0$, which yields $\Delta k \approx 2\sqrt{V_0 I}/(\hbar m)$. Hence, $k_\text{max} \gtrsim I\omega_L/(\hbar m) +  2\sqrt{V_0 I}/(\hbar m)$ is a reasonable choice. The diagonalization from Eq.~(\ref{eq: aes})  then yields $2 k_\text{max} + 1$ eigenstates. We further ensure that the lattice-trapped Bloch states are converged by comparing them with Bloch states obtained using basis sets with an over-sized $k_\text{max}$ ({\sl{i.e.}}, a $k_\text{max}$ larger than necessary). This ensures accuracy of the time-dependent simulations, explained next.

In the Sagnac PIC-TAILs, the lattice rotation rate is ramped up, held for some time, and ramped back down according to $\omega_L(t) = \pm \omega(t) + \Omega$, with a ramp function $\omega(t)$ that greatly exceeds the quasi-static platform rotation rate $\Omega$. The instantaneous Bloch states, $\vert \psi_l^n (t) \rangle$, and their energies, $E_l^n$ become time-dependent. Similar to Eq.~(\ref{eq: rseq}), we enter an arbitrary state $\ket{\psi(t)} = \sum_{n,l} d_l^n(t)\ket{\psi_l^n(t)}$ into the time-dependent Schr\"odinger equation for the Hamiltonian in Eq.~(\ref{eq: Htheta}), which yields 
\begin{align}\label{eq: aseq}
 \dot d_l^n(t) = \frac{E_l^n(t)}{i\hbar} d_l^n(t) - \sum_{n',l'} d_{l'}^{n'}(t) \braket{\psi_l^n(t)|\partial_t}{\psi_{l'}^{n'}(t)}.
\end{align}
Evaluation of the derivative term on the RHS requires the expansion coefficients $c_{l,k}^n (t)$ from Eq.~(\ref{eq: blochstate}), which are now also time-dependent.
Denoting the non-adiabatic coupling between the Bloch states
\begin{equation}\label{eq: hval}
 h_{l,n'}^{l,n}(t) = \sum_k (c_{l,k}^{n}(t))^* \frac{d}{dt} c_{l,k}^{n'}(t) \quad ,
\end{equation}
we get
\begin{align}\label{eq: aseqsimplify}
 \dot d_l^n(t) = \frac{E_l^n(t)}{i\hbar} d_l^n(t) - \sum_{n'} h_{l,n'}^{l,n}(t) \, d_{l}^{n'}(t) \quad .
\end{align}
There, the non-adiabatic couplings $h_{l,n'}^{l,n}$ rely on the time derivatives of the instantaneous Bloch basis states. The $h_{l,n'}^{l,n}$ are the manifestations of the classical Euler force from Eq.~(\ref{eq:ff}) in the described quantum formalism. The non-adiabatic couplings are non-zero only during the spin-up and spin-down stages of the TAI sequence, where $\omega_L$ depends on time. 

It is noted, again, that only states of equal quasi-momentum $l$ are coupled. The set of equations from Eq.~(\ref{eq: aseqsimplify}) is therefore block-diagonal in $l$, which enables reasonably fast integration for selected values of $l$. 
In practically relevant cases, the dynamics are nearly adiabatic, and we need only few Bloch states. Most of the computation time involves solving Eq.~(\ref{eq: aes}) to find the instantaneous Bloch-energies and -states on a sufficiently dense grid of $\omega_L(t)$-values. For fixed lattice depth $V_0$, this numerically intensive procedure needs to be executed only once; the results are stored in pre-calculated data sets. The eigen-values and eigen-states required when solving subsequently Eq.~(\ref{eq: aseqsimplify}) are then interpolated from these pre-calculated data sets.

The Schr\"odinger equation for the Hamiltonian in Eq.~(\ref{eq: Htheta}) can also be solved in the lattice-frame momentum-state basis, $\{ \ket{l,k} \}$. Although many more states are then required, there are still some advantages to that method. Firstly, solving the time-dependent equation does not require diagonalization of the Hamiltonian $\hat{H}_\theta$ on a dense grid of $\omega_L(t)$-values. Secondly, the representation of $\hat{H}_\theta$ in the momentum basis is a tri-diagonal matrix, while it is a dense matrix in the Bloch basis. If a backward numerical scheme is used, the complexity of Gauss elimination for a tri-diagonal matrix is $\mathcal{O}(N)$, instead of $\mathcal{O}(N^3)$ for a dense matrix of dimension $N$. Thirdly, in our simulations we include a significant number of free-particle bands to ensure convergence of the results. The free-particle energies typically undergo narrow anti-crossings during the $\omega_L$-sweeps. In momentum representation, the narrow anti-crossings do not require any special attention when numerically integrating the time-dependent Schr\"odinger equation for $\hat{H}_\theta$.
In contrast, in Bloch-state representation the anti-crossings entail rapid Bloch-state changes at times when the anti-crossings occur, leading to fast dynamics of 
the non-adiabatic coupling terms in Eq.~(\ref{eq: hval}). This feature requires careful attention when solving the time-dependent problem in the Bloch-state representation.

\section{Proposed PIC Design}
\label{sec:expt}

In this section, we analyze how a Sagnac TAI can be realized using PIC methods. $^{87}$Rb atoms are optically trapped in azimuthal optical lattices generated by evanescent light fields of optical micro-ring resonators on a photonic chip. Since the Sagnac sensitivity in Eq~(\ref{eq:sensitivity}) is proportional to the enclosed area $A$, we envision ring radii that are both feasible and not too large, as we will eventually seek low-SWaP implementations of a Sagnac PIC-TAI. Here, we will use $R_0 = \SI{600}{\um}$ as a specific example. The atoms are cooled down to a Bose-Einstein condensate (BEC) in an optical dipole trap, and then split and loaded onto the photonic chip, as described in some detail in Sec.~\ref{sec:concepts}. The interferometry will then be performed on the surface of the PIC. In this and the next section, we will focus on the chip details and on the atom dynamics that occur while the atoms are optically confined in the PIC-TAI. 

\begin{figure}[htb]
 \centering
 \includegraphics[width=0.45\textwidth]{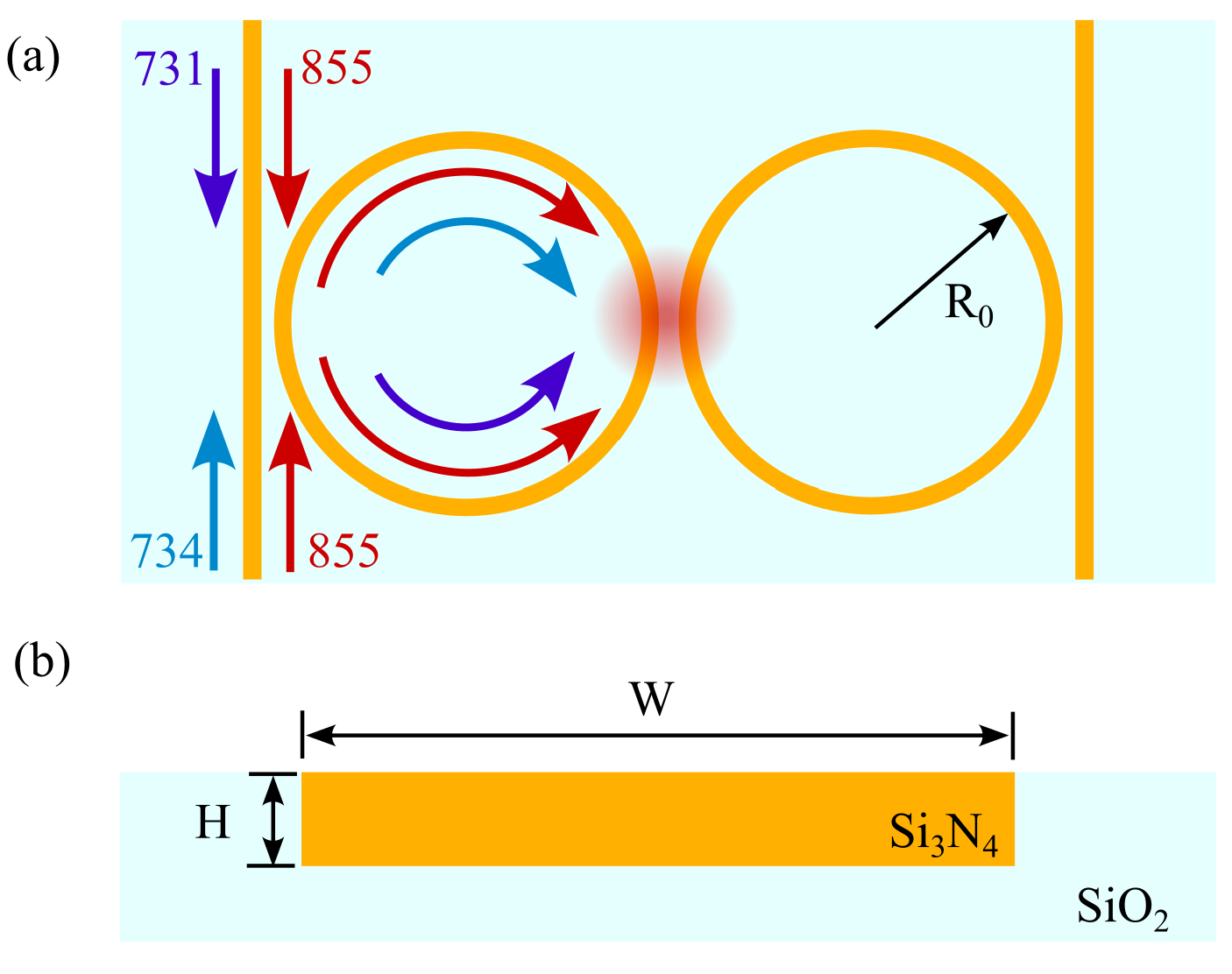}
 \caption{(a) Top-down view of the photonic chip (PIC). Two counterpropagating red-detuned (\SI{855}{nm}) modes form a trapping lattice above the ring. Two blue-detuned modes (\SI{734}{nm} and \SI{731}{nm}) repel the atoms from the waveguide. The red shaded area between two rings is the atom loading area. All wavelengths are vacuum wavelengths. (b) Cross-section through the optical waveguides. Parameters are $R_0 = \SI{600}{\um}$, $H = \SI{0.09}{\um}$, $W = \SI{1.8}{\um}$. The blue shading represents silicon dioxide (SiO$_2$), and the yellow is silicon nitride (Si$_3$N$_4$).}
 \label{fig:ring}
\end{figure}

The PIC structure is shown in Fig.~\ref{fig:ring}. The micro-ring resonators create combinations of running-wave and standing-wave circular optical lattices in evanescent optical fields. The waveguides that form the resonators and their feed lines are designed to support far-off-resonant TE00 field modes that are blue- or red-detuned  relative to the Rb D lines. The red-detuned TE00 modes will attract the atoms into their evanescent fields. Pairs of counter-propagating TE00 beams are injected into the feed lines and are proximity-coupled into cw and ccw TE00 field modes in the ring resonators. These in turn generate standing-wave azimuthal optical lattices above the resonators. In order to prevent the atoms from impacting onto the chip, and to keep attractive Casimir-Polder (CP) interactions between the atoms and the dielectric surface of the photonic chip at bay, running-wave blue-detuned TE00 modes are injected into the ring resonators to repel the atoms from the chip surface. The evanescent fields decay exponentially with $1/e$ decay distances of the intensity denoted $l_r$ and $l_b$ for the red and blue-detuned TE00 field modes, respectively.

Due to their differences in laser wavelength, the evanescent field of the blue-detuned TE00 mode decays faster than that of the red-detuned TE00 mode. As a result, if the light shift of the blue-detuned light at  $z = H/2$, the top surface of the waveguide, exceeds that of the red-detuned light, on the vertical symmetry plane of the waveguide, a potential minimum for the vertical atomic motion must exist above the chip surface. Neglecting the CP force, the position of the trap is determined only by the relative power ratio of the two colors, and the depth of the trap is determined by the absolute powers of the two colors.

The CP effect is ignored in the present analysis for the following two reasons. Firstly, the CP attraction is common-mode in the cw and ccw TAI rings. It is oriented vertically relative to the chip, and axially in the optical-lattice frame, so in lowest order it does not affect the dynamics of the interferometer. Secondly, we can tune the position of the trap to $\gtrsim \SI{200}{nm}$ above the surface, where the CP force becomes relatively small on the scale of the optical forces. We have estimated that for MHz-deep optical-lattice traps the CP force is $\lesssim 10\%$ of the maximum optical force of constraint, at $\sim \SI{200}{nm}$ above the surface. 

It is possible to create an optical-lattice atom trap with only fundamental TE00 modes of two colors~\cite{barnett2000substrate}. However, here we will excite a third, non-fundamental TE10 mode to increase tunability and to improve atom confinement in the plane parallel to the chip surface. The TE10 (rail) mode has a node at its center, $x = 0$, and it only has a small influence in the azimuthal and axial directions of the PIC rings. Since the waveguides confine shorter wavelengths of light more robustly than longer ones, it is not difficult to find a geometry that supports TE00 modes at red-detuned and blue-detuned wavelengths, while also supporting a higher-order TE10 blue-detuned rail mode. There is a weak restriction, however: the wavelengths of the two blue-detuned modes must sufficiently differ in frequency so that the beating of these two modes will not couple with the atomic motion.

\begin{figure}[htb]
 \centering
 \includegraphics[width=0.5\textwidth]{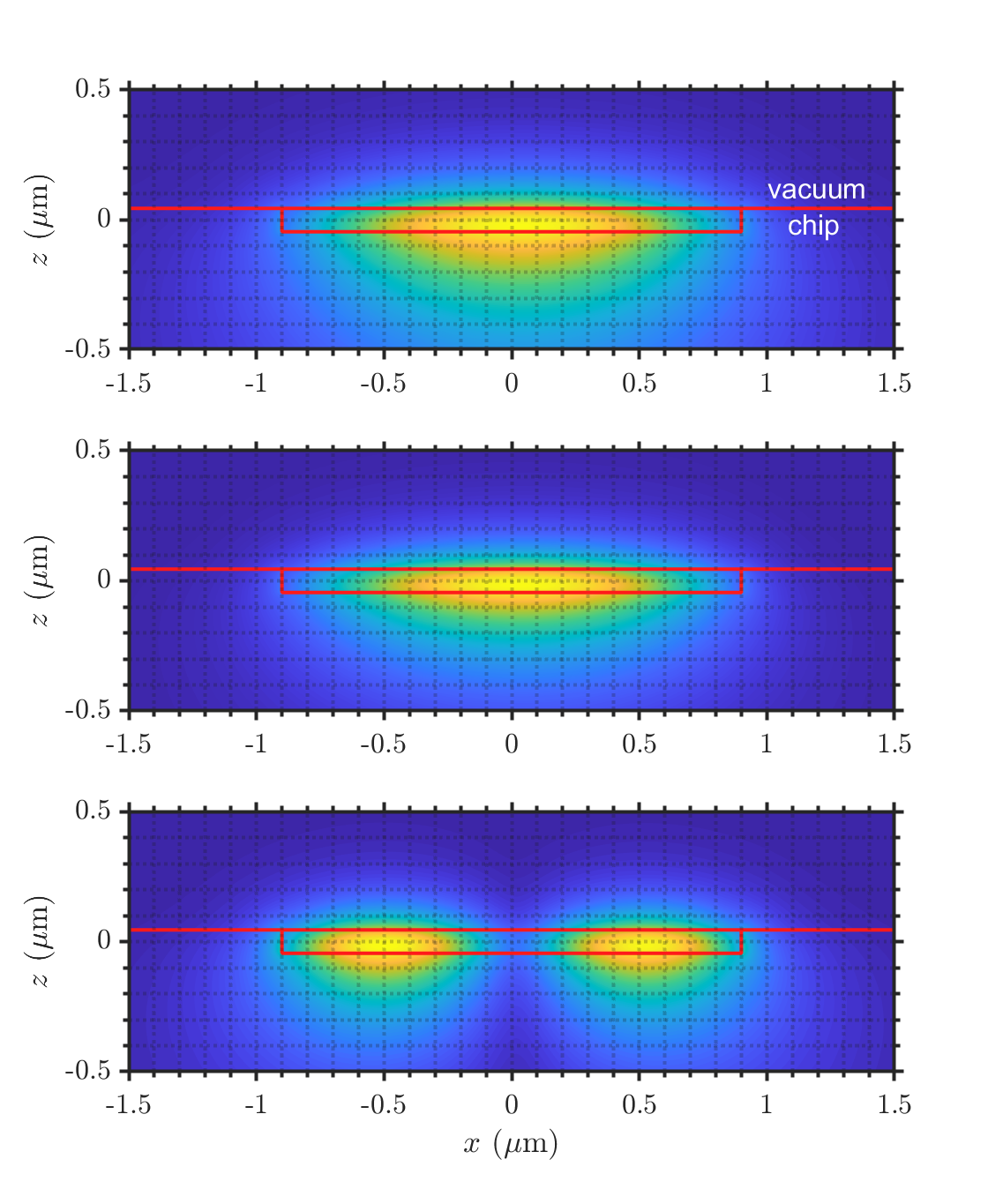}
 \caption{Electric field magnitudes of the $\SI{855}{nm}$ TE00, $\SI{731}{nm}$ TE00, and $\SI{734}{nm}$ TE10 modes, from top to bottom. The  red rectangle outlines the Si$_3$N$_4$ core of the waveguide. All wavelengths are vacuum wavelengths.}
 \label{fig:waveguide_modes}
\end{figure}

\begin{figure}[htb]
 \centering
 \includegraphics[width=0.5\textwidth]{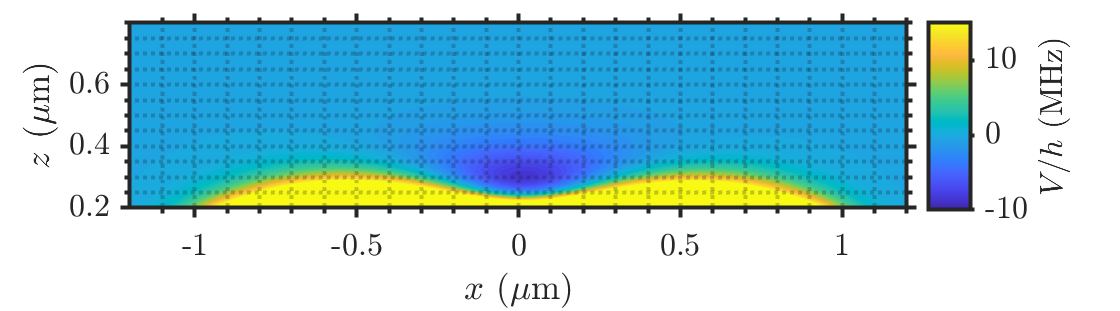}
 \caption{Trapping potential for Rb $5S_{1/2}$ atoms along the radial and axial directions of the PIC rings at an antinode of the azimuthal PIC-TAIL. The intra-cavity circulating powers of the $\SI{855}{nm}$ TE00 azimuthal-lattice modes, the $\SI{731}{nm}$ TE00 running-wave mode, and the $\SI{734}{nm}$ TE10 running-wave mode are $\SI{75}{mW}$ (each lattice mode), $\SI{400}{mW}$, and $\SI{200}{mW}$, respectively. The potential minimum has a value of $-\SI{8.3}{MHz}$ and is located at $z = \SI{0.310}{\um}$, which is $\SI{0.265}{\um}$ above the surface of the PIC.}
 \label{fig:potential_cross_section}
\end{figure}

A PIC ring-resonator can be designed with commercially available software, such as COMSOL. Given a definite structure, like the one shown in Fig.~\ref{fig:ring}, COMSOL is able to find its resonating modes. We have modeled a waveguide of a Si$_3$N$_4$ core with dimensions shown in Fig.~\ref{fig:ring}, built on SiO\textsubscript{2}. We exploit the cylindrical symmetry of the ring resonator to reduce computation time. The geometry is single-mode (TE00) at wavelength $\sim\SI{855}{nm}$, while also supporting higher-order modes at $\sim\SI{730}{nm}$. The fields of a red-detuned TE00-mode at $\SI{855}{nm}$, a blue-detuned TE00-mode at $\SI{731}{nm}$, and a blue-detuned TE10-mode at $\SI{734}{nm}$ are displayed in Fig.~\ref{fig:waveguide_modes}. 

The potentials experienced by the atoms are the sum of the light shift potentials induced by the three individual field colors,
\begin{align}\label{eq: lightshift}
 V = - \sum_{f_i} \frac{1}{4}\alpha(f_i) |\mathcal{E}_{f_{i}}|^2
\end{align}
where the $\alpha(f_i)$ are the ac electric polarizabilities, which are scalar for Rb $5S_{1/2}$ states, the $f_i$ the three laser colors (855~nm, 731~nm and 734~nm), and the $\mathcal{E}_{f_{i}}$ the electric-field amplitudes at the field frequencies $f_i$. Notably, the fields of the two counter-propagating 855-nm modes that form the PIC-TAILs have to be summed coherently before taking the square in Eq.~(\ref{eq: lightshift}). Fig.~\ref{fig:potential_cross_section} shows the cross-section of the trapping potential at a lattice anti-node for a sample set of circulating powers of the four ring-resonator modes. The figure shows tight trapping in the axial and radial directions. 

\begin{figure}[htb]
 \centering
 \includegraphics[width=0.5\textwidth]{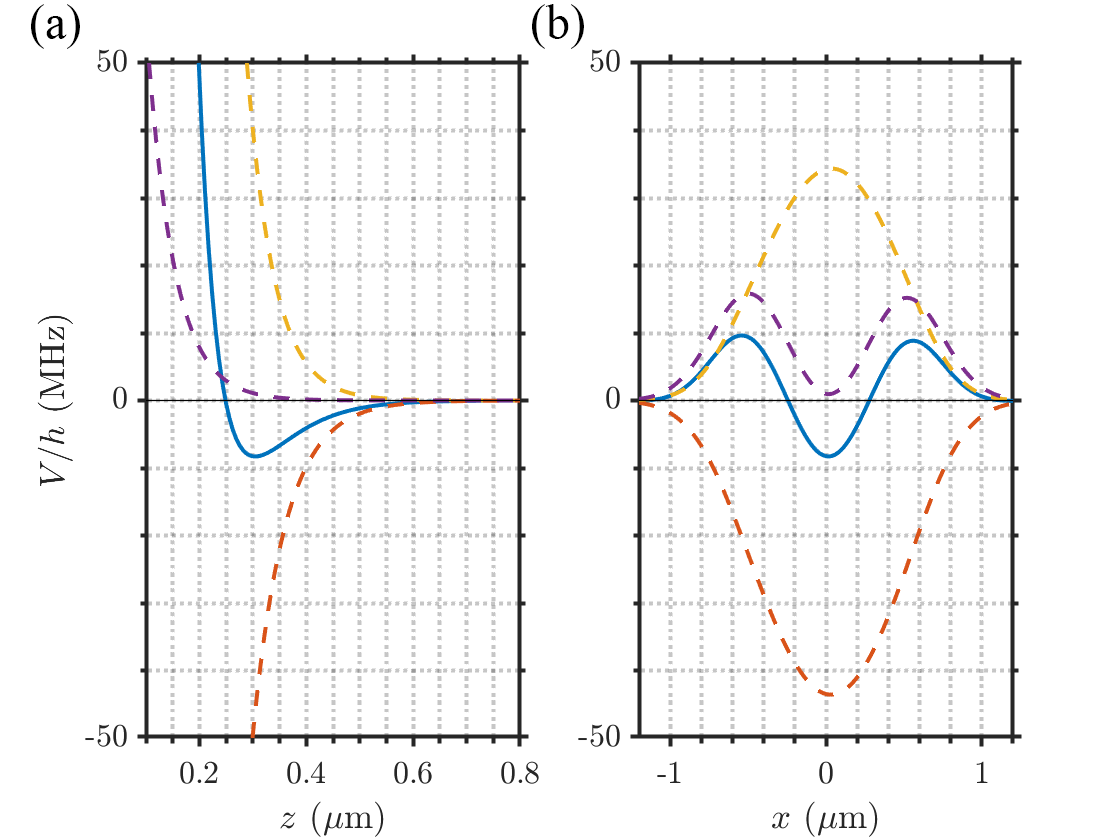}
 \caption{(a) Trapping potential transverse to the chip surface, at $x = 0$ of Fig.~ \ref{fig:potential_cross_section}. (b) Trapping potential parallel to the chip surface and transverse to the direction of the PIC waveguides at $z = \SI{0.31}{\um}$. The red, yellow, and purple dashed curves show the contributions to the trapping potential from the $\SI{855}{nm}$ TE00, $\SI{731}{nm}$ TE00, and $\SI{734}{nm}$ TE10 modes, respectively, for powers as in Fig.~ \ref{fig:potential_cross_section}. The blue curve is the sum of all modes.}
 \label{fig:potential_axial}
\end{figure}

For the conditions of Fig.~\ref{fig:potential_cross_section}, the characteristic decay lengths of the intensities of the evanescent TE00 waves at $x = 0$ are $l_r = \SI{124}{nm}$ and $l_b = \SI{101}{nm}$. Here, it is noted that a large difference between the two characteristic lengths is crucial for effective trapping transverse to the PIC surface. The difference in characteristic lengths can be optimized with the geometry of the waveguide, and by increasing the frequency difference between the red-detuned and the blue-detuned TE00 modes. The trapping behavior transverse to the chip surface, for our exemplary case, is illustrated by Fig.~\ref{fig:potential_axial}~(a). In Fig.~\ref{fig:potential_axial}~(b) it is seen that the TE01 mode is critical for providing tight trapping in the direction parallel to the chip surface and transverse to the guide. 

\begin{figure}[htb]
 \centering
 \includegraphics[width=0.5\textwidth]{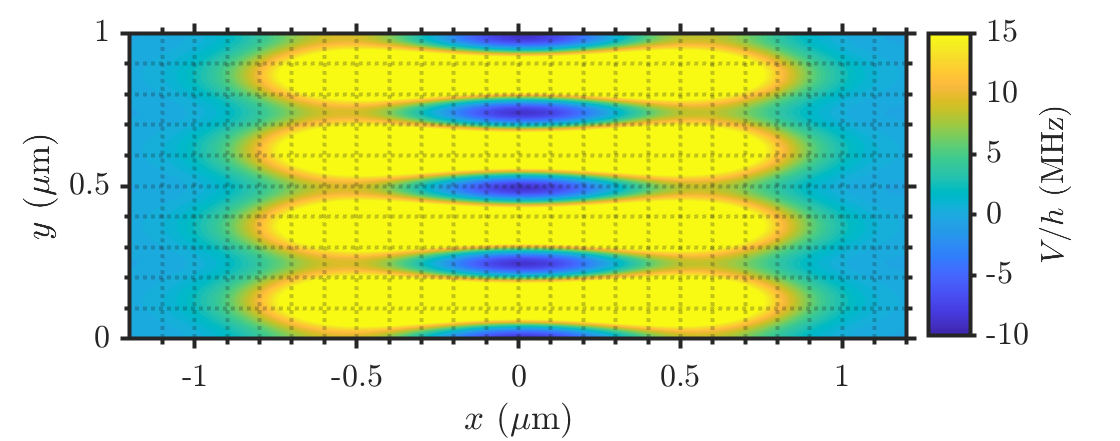}
 \caption{Radial and azimuthal atom-trapping potential on a plane containing the potential minimum, for powers as in Fig.~ \ref{fig:potential_cross_section}. The vertical axis in the plot is $y = R_0\theta$, 
 and the horizontal axis points in the radial direction of the PIC ring.}
 \label{fig:potential_xy}
\end{figure}

Fig. \ref{fig:potential_xy} shows a top-down view of the lattice trapping potential on the plane that passes trough the potential minima. The figure illustrates the structure of the PIC in the the azimuthal and radial directions of the PIC rings. Azimuthal trapping is relatively tight in comparison with radial trapping. The absolute lattice depth is tunable by adjusting the powers in the waveguide modes.

We next describe how the PIC-TAIL is spun up and back down to effectuate a Sagnac AI.  The lattice is made of two counter-propagating beams of the same transverse TE00 mode and near-identical frequencies $\omega_0 \pm \Delta \omega(t)/2$, where $\omega_0$ coincides with a resonant PIC-resonator frequency near $c/\SI{855}{nm}$. This mode generates the attractive azimuthal optical-lattice potential that pulls the atoms towards the chip surface. $\Delta \omega (t)$ denotes a time-dependent frequency difference between the counter-propagating lattice beams. The lattice beams are injected into the rings by corresponding TE00 field modes that counter-propagate through the straight coupler-guide sections, as shown in Fig.~\ref{fig:ring}; the beams are electro-optically derived from a common laser beam, using (anti-)symmetric high-fidelity RF-modulator ramps to minimize differential phase noise on these beams. The 855-nm TE00 modes in the Si$_3$N$_4$ guides have a group index $n_g = 1.685$, given by the free spectral range of $\SI{47.19}{GHz}$ found in the simulation, and a phase index $n_p = 1.474$, given by the number of lattice intensity maxima along the ring, which is $m = 13000$. The PIC-TAILs in the two resonator rings then counter-rotate, in the instrument frame, at angular frequencies $\omega_L(t) = \pm \Delta \omega / (2 k R_0)$, where $k = 2 \pi n_p / \lambda$ is the propagation wave number of the $\lambda=855$-nm TE00 modes. The maximum  $|\omega_L|$ is limited by the linewidth of the resonator resonances, $\kappa$. Hence, there is a trade-off between the resonator power enhancement and the maximum spin rate, $|\omega_L|$, of the PIC-TAILs. The lower the $Q$-factor, the higher the maximum $|\omega_L|$, but the more power we need to create the PIC-TAILs. Assuming a $\SI{1.5}{dB/m}$ loss of the 855-nm lattice modes waveguides, corresponding to an intrinsic $Q$-factor of $\approx 3.6\times 10^7$, and assuming then a coupling-limited $Q$-factor of $\sim 5\times 10^6$, the cavity-mode linewidth is $\kappa = \omega/Q \approx 2\pi\times\SI{70}{MHz}$. The splitting $\Delta \omega$ required for an angular frequency $\omega_L$ of the azimuthal rotation of the PIC-TAILs is $\Delta \omega = \omega_L 4 \pi R_0 n_p / \lambda$. Requiring that both lattice modes at optical angular frequencies $\omega_0 \pm \Delta \omega(t)/2$ must fit under the same cavity resonance, one finds an upper limit $|\omega_L| \lesssim 2 \pi \times \SI{5}{kHz}$, which is well above the maximum $|\omega_L|$ of $2 \pi \times \SI{1}{kHz}$ used in simulations presented in Sec.~\ref{sec:sims}. Furthermore, the power gain for PIC-TAIL resonators with the stated coupling-limited $Q$-factor is $\sim 370$, so that the interferometer would only require a few mW of optical power to form PIC-TAILs with atom-trapping depths of $V_0 \sim h \times \SI{10}{MHz}$.

To account for any manufacturing asymmetries of the cw and ccw-rotating PIC-TAILs, the respective applied RF ramps may need to be fine-tuned to synchronize the azimuthal phases in the instrument frame at a precision of a few mrad over the duration of the Sagnac TAI loops. For $K=1000$ and a number of $m$ lattice wells on the rings, this translates into a relative RF stability requirement for the RF sources used in the range of a few $10^{-3} / (2 \pi K m) \sim 10^{-10}$. Such sources are commercially available.

\section{Simulation}
\label{sec:sims}

In this section, we numerically explore non-adiabatic effects, adiabaticity, and sensitivity of the exemplary Sagnac PIC-TAI discussed in Sec.~\ref{sec:expt}. Sensitivity is largely derived from a large number of full rotations, $K/2$, per measurement sequence. As such, the PIC-TAILs must be spun up over a short time $t_r$ through the above defined ramp function, $\omega(t)$, to a large terminal value denoted $\omega_s$, held at terminal speed for a (usually much longer) time $t_c$, and back down with a reverse ramp function of duration $t_r$, for eventual TAI readout. The PIC-TAILs rotate at $\omega_L(t) = \pm \omega(t) + \Omega$. We compare linear ramp-up functions, $\omega(t) = \omega_s \, t/t_r$, and smooth sinusoidal ramp functions, $\omega(t) = \omega_s  \, \sin^2(\pi t/2t_r)$, with $0 \le t \le t_r$. The ramp-downs occur in an analogous reverse manner. The instrument rotation to be measured, $\Omega \ll \omega_s$, is assumed to be fixed during one measurement cycle. If $\Omega$ was a slow function of time, we would measure the average of $\Omega$ over a cycle. In our present model, radial and azimuthal dynamics are independent, implying that the Coriolis force has no effect (as is our assumption).

\subsection{Radial Dynamics Simulation}

The example shown in Sec.~\ref{sec:expt} has a radial potential of full depth $\Delta V = h \times \SI{17}{MHz}$ and a trap frequency of $\omega_r = 2\pi \times \SI{183}{kHz}$ near the bottom. The maximum spin rate the potential can support is $\omega_{s,\max} \approx 2\pi\times \SI{3.2}{kHz}$, which is found by equating the maximum radial trapping force with the centrifugal force for $R_0 = \SI{600}{\um}$. We choose $\omega_s = 2\pi\times \SI{1}{kHz}$, where the radial force is about one-tenth of the maximum supported. The PIC-TAIL frequencies of the pair of 855-nm TE00-modes required for this $\omega_s$-value differ by $<\SI{15}{MHz}$ and fit well within the cavity linewidth calculated in Sec.~\ref{sec:expt}. The full trap depth at that $\omega_s$ is $\Delta V = h\times \SI{14}{MHz}$, and the number of radially bound states $N \gtrsim \Delta V/ (\hbar\omega_r) \approx  80$. Here, it is sufficient to use $N=20$ basis states because we are only interested in cases that are near the adiabatic limit.

\begin{figure}[htb]
 \centering
 \includegraphics[width=0.5\textwidth]{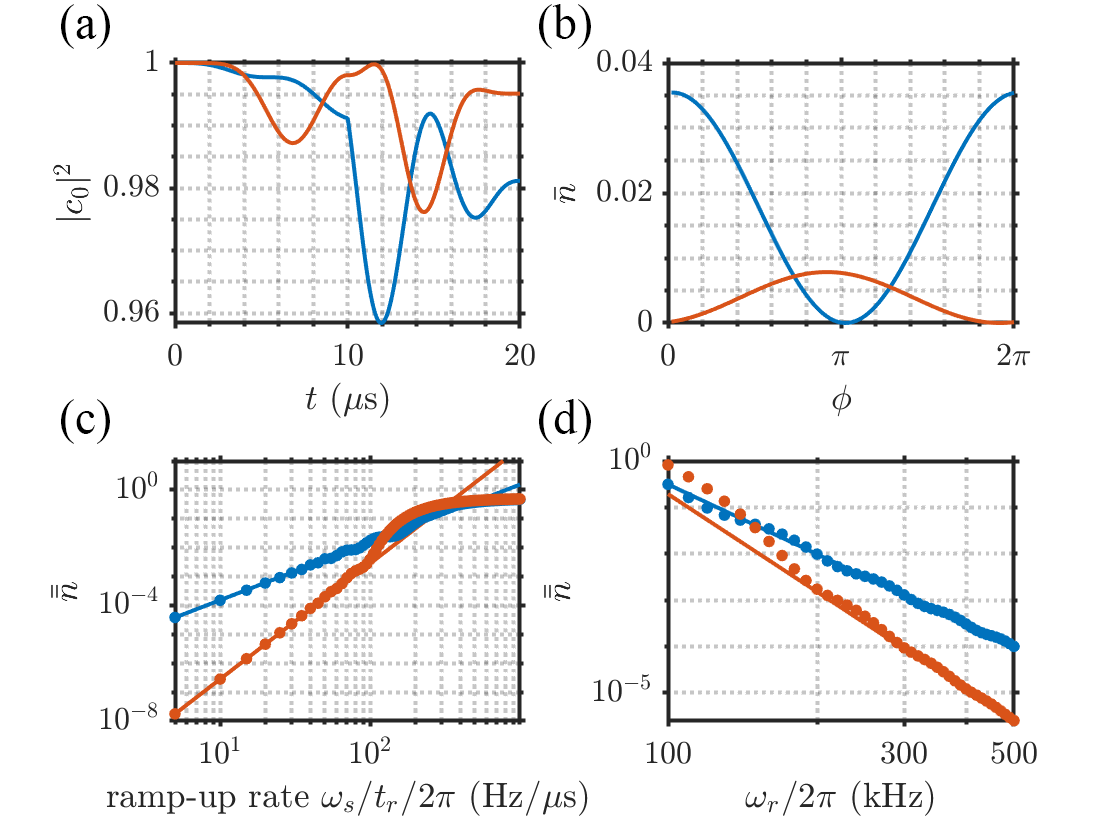}
 \caption{(a) Radial ground-state population during the PIC-TAI sequence described in the text. The maximum spin rate is $\omega_s = 2\pi\times \SI{1}{kHz}$, the ramp time $t_r = \SI{10}{\us}$, the radial trap frequency $\omega_r = 2 \pi \times \SI{183}{kHz}$. For the specific example displayed,  the phase acquired in the constant rotation stage is arbitrarily set to $\phi = \pi/2$. In all four panels, the blue curves and symbols represent linear ramps, and the red curves and symbols $\sin^2$ ramps. (b) Final-state expectation value $\bar n$ vs phase $\phi$, with other parameters as in (a). (c) $\phi$-average, $\bar{\bar n}$, vs $\omega_s/t_r$ for fixed $\omega_s = 2\pi\times \SI{1}{kHz}$ and $\omega_r = 2 \pi \times 183$~kHz. (d) $\bar{\bar n}$ vs $\omega_r$, for $\omega_s = 2\pi\times \SI{1}{kHz}$ and  $t_r = \SI{10}{\us}$. The symbols are simulated data points, and the curves are analytical solutions from Eqs.~(\ref{eq: nbarstaradlimit}) and~(\ref{eq: nbarstaradlimitsin2}).}
 \label{fig:linear_sin2_compare}
\end{figure}

The radial simulation is performed in the adiabatic basis of Sec.~\ref{sec:dynamics}~B. Non-adiabatic radial state mixing occurs during the ramps [see Eq.~(\ref{eq: rseqsimplify})]. During the constant-rotation stage the Hamiltonian is diagonal, and the wave-function coefficients $c_n$ merely pick up phases according to $c_n(t_r) \longrightarrow c_n(t_r)\exp(-i n \phi)$, where $\phi = \omega_r t_c$. An example for the ground-state population evolution over one PIC-TAI sequence is shown in Fig.~\ref{fig:linear_sin2_compare}~(a). The kinks at $t=t_r= \SI{10}{\us}$ occur due to said phase jumps, which are applied at time $t_r$ to include a constant-rotation stage in the simulation. The decrease in the ground-state population reflects the non-adiabatic effects during spin-up and -down. We use the final-state expectation value of the number operator, $\bar n = \sum_{n = 0}^\infty |c_n|^2 n$, as our metric of non-adiabaticity.

The final state is sensitive to the phase jump $\phi$. Fig.~\ref{fig:linear_sin2_compare}~(b) shows that $\bar n$ has a sinusoidal dependence on $\phi$. As an average metric for non-adiabaticity, we define the $\phi$-average
\begin{align}\label{eq: nmean}
 \bar {\bar n} = \frac{1}{2\pi}\int_0^{2\pi} \bar n(\phi)\ d\phi \quad.
\end{align}
This metric of non-adiabaticity eliminates the dependence on $\phi$. The dependence of $\bar{\bar n}$ on the average ramp rate $\omega_s / t_r$ is shown in Fig.~\ref{fig:linear_sin2_compare}~(c). Over most of the range, the non-adiabaticity increases according to a power-law in $\omega_s / t_r$. As expected, in the lower $\omega_s / t_r$-range, where the PIC-TAI should be operated, the $\sin^2$ ramps result in better fidelity ({\sl{i.e.,}} lower $\bar{ \bar n}$) than linear ramps.  For instance, to keep $\bar{\bar n}$ below $1\%$, $\omega_s / t_r$ has to be smaller than $\SI{85}{Hz/\us}$ or $\SI{110}{Hz/\us}$, corresponding to $t_r > \SI{12}{\us}$ or $t_r > \SI{9}{\us}$, for linear and $\sin^2$ ramps, respectively. In the case shown in Fig.~\ref{fig:linear_sin2_compare}~(c), the curves cross at $\omega_s/t_r = \SI{125}{Hz/\us}$, where $\bar{\bar n}= 0.024$, as the system transitions into a quantum projection regime in which the difference in behavior between linear and $\sin^2$-ramps disappears (see below).

To understand the power scaling apparent in Fig.~\ref{fig:linear_sin2_compare}~(c), we analytically solve the radial dynamics in the small-$\dot\omega$ limit in first order perturbation theory. We consider only the transition from the ground state to the first excited state,
\begin{align}\label{eq: firststate}
 \dot c_1 = -i\omega_r c_1 - \sqrt{\frac{M\omega_r}{2\hbar}} \frac{d}{dt} \delta R
\end{align}
with the initial condition $c_1(0) = 0$ and $d \delta R /dt$ from Eq.~(\ref{eq: x0dot}). The solution is
\begin{align}\label{eq: firststatesol}
 c_1(t) = - \sqrt{\frac{M\omega_r}{2\hbar}} \exp(-i\omega_r t)\int_0^t \left[ \frac{d}{dt'} \delta R(t') \right] \exp(i\omega_r t') \ dt' \quad.
\end{align}
Since only the first excited state is considered, it is $\bar n \approx |c_1|^2$. For the linear ramp, Eq.~(\ref{eq: firststatesol}) yields
\begin{align}\label{eq: firststatesollinear}
 \bar n_\text{lin} = 8\frac{M R_0^2}{\hbar}\frac{\dot\omega^4}{\omega_r^7} |u|^2 \sin^2(\xi - \omega_r t_f/2)
\end{align}
where $u = |u|\exp(i\xi) = (1-i\omega_r t_r)\exp(i\omega_r t_r) -1$, and $t_f = 2t_r + t_c$ is the final time. In the perturbative limit,  $\omega_r t_r \gg 1$, and hence $|u| = \omega_r t_r$.

The argument of the $\sin^2$-term in Eq.~(\ref{eq: firststatesollinear}) includes the phase accumulation from the constant-rotation stage, $\phi = \omega_r t_c$, which manifests as a sinusoidal dependence of $\bar{n}$ on $t_c$. This sinusoidal dependence is also evident in the simulation result in Fig.~\ref{fig:linear_sin2_compare}~(b). The physical reason behind the sinusoidal dependence is that the spin-up and -down ramps are symmetric and produce the same amount of (first-order) complex excitation amplitudes, with the time interval between them inducing a phase shift of the second excitation relative to the first, akin to a Ramsey interference experiment. Averaging over the phase shift, we find
\begin{equation}\label{eq: nbarstaradlimit}
 \bar {\bar n}_\text{lin} = 4\frac{MR_0^2}{\hbar}\frac{\dot\omega^4}{\omega_r^7} |u|^2 
 = 4\frac{MR_0^2 \, \omega_s^2}{\hbar}\frac{\dot{\omega}^2}{\omega_r^5}.
\end{equation}
We see that for the linear ramp $\bar {\bar{n}}_\text{lin}$ is proportional to $\dot\omega^2 = (\omega_s / t_r)^2$ and $\omega_r^{-5}$, as reflected by the red simulation data (symbols) and analytical results (lines) in Figs.~\ref{fig:linear_sin2_compare}~(c) and~(d).

Integrating Eq.~(\ref{eq: firststatesol}) for the case of our $\sin^2$-ramps yields
\begin{align}\label{eq: nbarstaradlimitsin2}
 \bar {\bar{n}}_\text{sin} &= \pi^4\frac{MR_0^2}{\hbar}\left(\frac{\omega_s}{t_r} \right)^4 \frac{1}{\omega_r^7} \quad.
\end{align}
Hence, for the sin$^2$-ramps, $\bar {\bar n}_\text{sin}$ is proportional to $(\omega_s / t_r)^4$ and $\omega_r^{-7}$, in agreement with the blue simulation data (symbols) and analytical results (lines) in Figs.~\ref{fig:linear_sin2_compare}~(c) and~(d). Our analytical analysis  therefore confirms that in the adiabatic regime, $\sin^2$-ramps scheme are considerably better than linear ones.

The non-adiabatic, fast-ramp limit of $ \bar {\bar n} \sim 0.5$ seen in Fig. ~\ref{fig:linear_sin2_compare}~(c) also has an analytical explanation. In the fast-ramp limit the spin-up and -down ramps take the form of sudden changes of the Hamiltonian, regardless of ramp type. The ramp-up suddenly translates the ground state inward (relative to the outward-shifting radial potential) by a distance $\delta R$, {\sl{i.e.}} after the ramp-up the vibrational state in the radial potential's frame is a coherent state, $\ket{\psi(t_r)} = \hat{T}(-\delta R)\ket{0} = \ket{-\alpha_0}$, with $\alpha_0 = \delta R /\sqrt{2 \hbar / (M \omega_r)}$. The time evolution due to the fixed-rotation stage rotates the coherent state, so that $\ket{\psi(t_r+t_c)} = \ket{-\exp(-i\phi) \alpha_0}$, where $\phi = \omega_r t_c$. The spin-down stage translates the state a second time, namely outward by $\delta R$, and the final state at time $2 t_r + t_c$ is a coherent state 
\begin{equation}\label{eq: rdfs} 
 \ket{\psi_f} = \hat{T}(\delta R)\ket{\psi(t_r + t_c)} 
 = \ket{\alpha_0 - \exp(-i\phi) \alpha_0} \quad.
\end{equation}
The average excitation number after the second sudden translation then is 
\begin{equation}\label{eq: rdfsnbar}
 {\bar {n}_\text{non-ad}} = |\alpha_0 - \exp(-i\phi) \alpha_0|^2 \\
 = 4 \alpha_0^2 \sin^2\frac{\phi}{2}
\end{equation}
and, averaging over $\phi$,
\begin{align}\label{eq: rdfsnbarstar}
 \bar {\bar n}_\text{non-ad} &= 2\alpha_0^2.
\end{align}
In our case with $\omega_r = 2 \pi \times \SI{183}{kHz}$, $\omega_s = 2 \pi\times \SI{1}{kHz}$ and $R_0=600~\mu$m, we have $\delta R = \SI{18}{nm}$, $\alpha_0 = \delta R/\sqrt{2 \hbar / (M \omega_r)}= 0.50$, and $\bar{\bar n} = 0.50$ in the non-adiabatic limit, in agreement with Fig.~\ref{fig:linear_sin2_compare}~(c). In the Sagnac PIC-TAI application, the non-adiabatic limit is of no practical importance.

\subsection{Azimuthal Dynamics Simulation}

\begin{figure}[htb]
 \centering
 \includegraphics[width=0.5\textwidth]{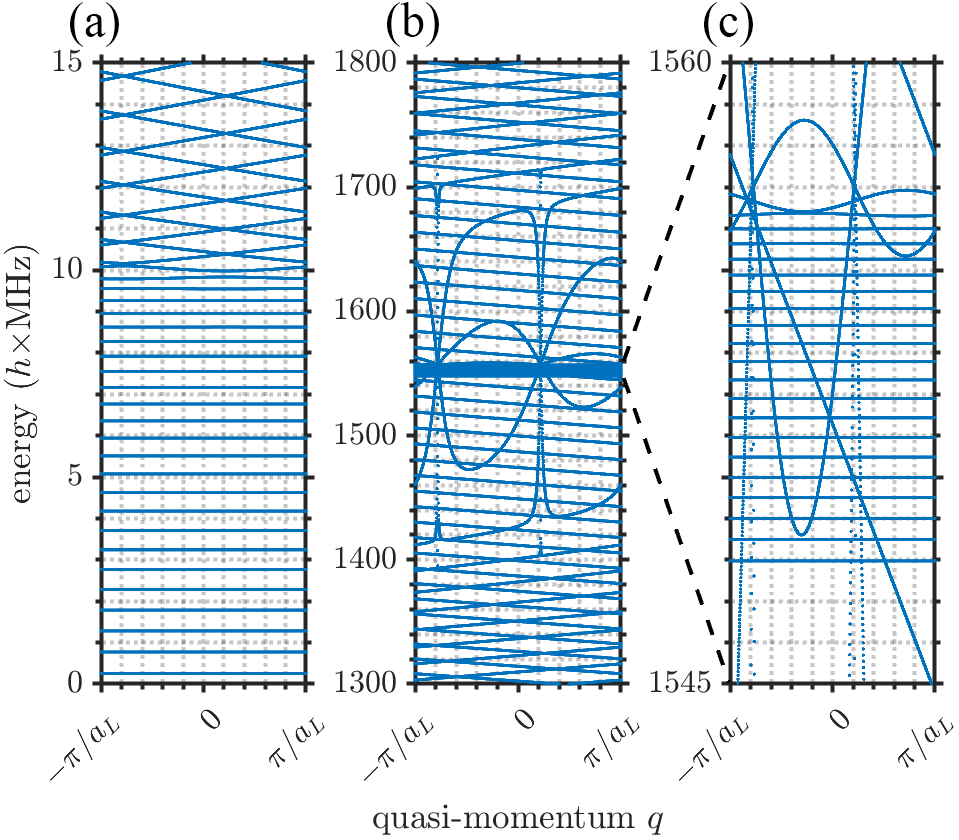}
 \caption{Band structure in a lattice with full depth $2V_0 = h \times \SI{10}{MHz}$ and spin rate $\omega = 2\pi\times \SI{1}{kHz}$. The quasi-momentum $q$ is discrete and is related to the integer quantum number $l$ in Sec.~\ref{sec:dynamics}~C  via $l=qR_0$. (a) Band structure in the co-rotating frame. While at energies above $h \times 10$~MHz it is asymmetric with respect to $q = 0$, the tightly bound bands are essentially identical to those in a non-rotating lattice. (b) Band structure in the instrument frame for $\Omega=0$, and (c) expanded view of (b) in the energy range of the lattice-trapped bands. The lattice-trapped bands in the instrument frame bottom out at an energy equal to the kinetic energy of an atom that co-rotates with the lattice, $I\omega_L^2/2  \approx h\times\SI{1548}{MHz}$.}
 \label{fig:band}
\end{figure}

In Fig.~\ref{fig:band} we show the band structure for the azimuthal potential for a PIC-TAIL with full lattice depth $2 V_0 = h \times \SI{10}{MHz}$. The lattice has a number of $m$=13000 lattice wells along the $R_0 = \SI{600}{\um}$ rings, corresponding to a lattice constant of $a_L \approx \SI{290}{nm}$, or half the wavelength of the 855-nm TE00 modes inside the waveguide.

The atomic band structure in the co-rotating frame is shown in Fig.~\ref{fig:band}~(a). The bottom of the lattice can be approximated as a harmonic potential with frequency $\omega_{\theta} = \sqrt{\frac{V_0}{M}}\frac{2\pi}{a} \approx 2\pi\times \SI{0.5}{MHz}$, which is also the spacing of the trapped bands. Near the top of the lattice wells at $h \times \SI{10}{MHz}$, the bands transition from trapped into free-particle bands, which are asymmetric for non-zero $\omega_L$ [as in Fig.~\ref{fig:band}~(a)]. The band structure in the instrument frame is shown in Figs.~\ref{fig:band}~(b) and~(c). The instrument-frame band structure with the lattice turned off consists of free-particle bands, which have the shape of intersecting sections of parabolas. With the lattice on, a number of bands with velocities near the rotating-lattice velocity collapse into an array of tightly bound bands near the kinetic energy $I\omega_L^2/2 \approx h\times\SI{1548}{MHz}$. Bands that are far from that energy, as well as bands with velocities opposite to  the lattice rotation direction, largely remain free-particle bands. Figure~\ref{fig:band}~(a) and the zoom in Fig.~\ref{fig:band}~(c) allow a detailed comparison between the lattice-trapped bands, which are the most relevant to our work, in the lattice and the instrument frames. In our simulation, we initialize the atoms in the $l = 0$ state of the lowest band. In position representation, the atoms have an equal probability to be in the ground state of any lattice site. While this state is globally coherent, well-to-well coherence is not required for the Sagnac-TAI principle of operation simulated next. As such, the quasi-momentum quantum number $l$ is not critical. 

The atom interferometer consists of two lattices, which have angular velocities $\omega_L(t) = \pm \omega(t) + \Omega$. We simultaneously simulate both lattices to extract the interferometric phase difference. For simplicity, we assume that the cw- and ccw-rotating PIC-TAILs are overlapped in space and are acting selectively on two different internal spin states of the atom. Initially, the lattices are at rest in the instrument frame, and the atoms are initialized in one of the two spin states with $l=0$, in the ground band of the lattice. We then apply an instantaneous Hadamard operation between the spin-up and -down states of the internal atomic spin. This operation prepares the atoms in symmetric superpositions of wave-packet states, split between the vibrational ground states in overlapping wells of the two spin-selective lattices. Coherence between wave-packet components in initially overlapping pairs of wells of the cw- and ccw-rotating PIC-TAILs is required, but coherence between neighboring wells in any of the two PIC-TAILs is not required. The two lattices are then spun up in a cw and ccw manner, following $\omega_L(t) = \pm \omega(t) + \Omega$, and the Schr\"odinger equations are solved independently for the two lattices. We spin up the lattices to $\omega_s = 2\pi\times \SI{1}{kHz}$, the peak value of $\omega(t)$, over a ramp-up time of $t_r = \SI{100}{\us}$. The rotation is then kept constant at $\omega_s$ for a time $t_c$, and then the lattices are ramped back down (over the same time $t_r = \SI{100}{\us}$). The ramps induce negligible fidelity loss from radial non-adiabaticity (see Sec.~\ref{sec:sims}~A). Both linear and $\sin^2$ ramps are simulated. In the end, the atomic wave-packets are recombined with a second Hadamard matrix. Accurate azimuthal phase matching between the cw and ccw rotation angles at recombination ensures that initially coherently-split wave-packet components are brought back together (also see last paragraph in Sec.~\ref{sec:expt}). The second Hadamard operation then leads to atom interference, at which point we measure the total number of atoms in each spin state, $N_\uparrow$ and $N_\downarrow$. This could be accomplished by spin-selective fluorescence readout. We finally calculate their ratio to get the quantum AI phase,
\begin{align}\label{eq: ratio}
 \cos^2(\Delta\phi_Q/2) = \frac{N_\uparrow}{N_\uparrow + N_\downarrow}.
\end{align}

\begin{figure}[htb]
 \centering
 \includegraphics[width=0.5\textwidth]{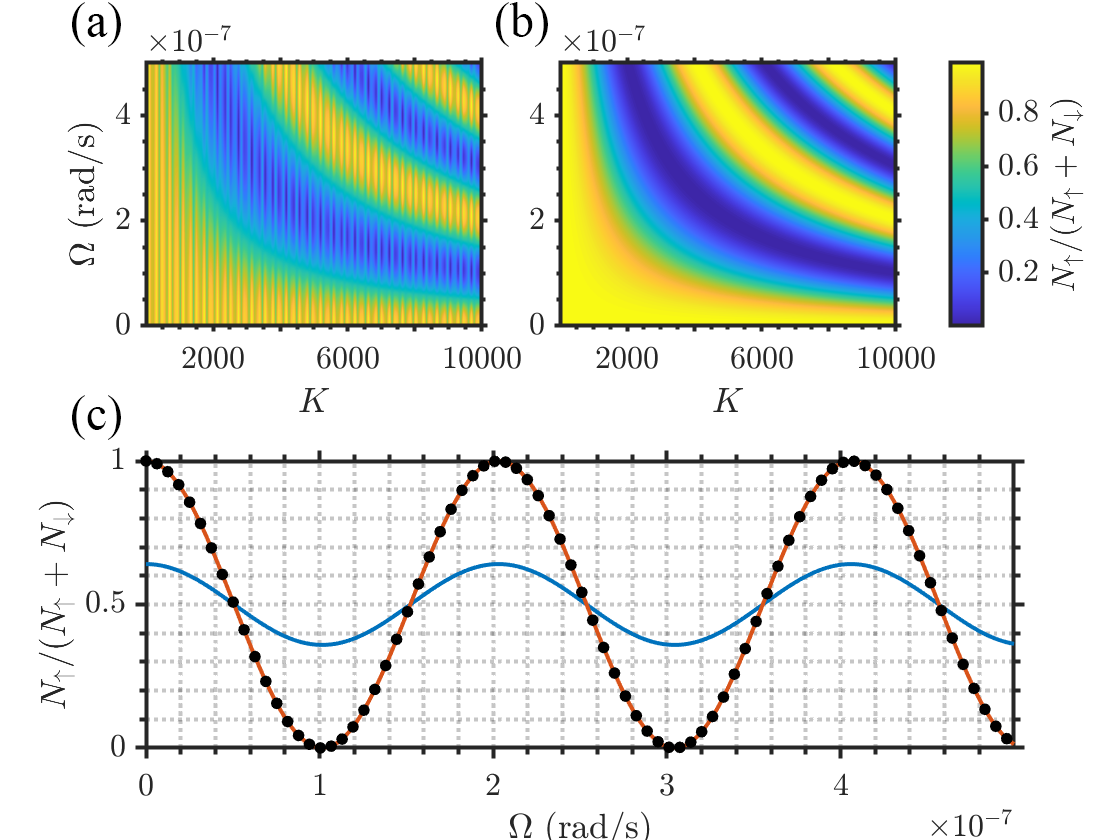}
 \caption{Simulated interferometric signal, Eq.~(\ref{eq: ratio}), over a range of angular velocities of the instrument rotation, $\Omega$, and number of half-turns, $K$. In panel (a), the ramp control function $\omega(t)$ is ramped up linearly, while in (b) it has a smooth $\sin^2$-shape. (c) AI signal vs $\Omega$ for $K = 10000$. The blue curve represents the linear-ramp and the red the $\sin^2$-ramp case. The black dotted line shows the semiclassical result according to Eq.~(\ref{eq:semiclassical}).}
 \label{fig:map}
\end{figure}

In Figs.~\ref{fig:map}~(a) and~(b) we show simulation results for the interferometric signal, Eq.~(\ref{eq: ratio}), over various values for angular velocity $\Omega$ and number of half-turns $K$ for linear and $\sin^2$ ramps, respectively. The data for linear ramps [Fig.~\ref{fig:map}~(a)] exhibit vertical stripes, which result from non-adiabaticity of the azimuthal motion. The data for sin$^2$ ramps [Fig.~\ref{fig:map}~(b)] are virtually free of manifestations of non-adiabaticity. At fixed $K$, the non-adiabaticity reduces the visibility of the interferometric signal versus $\Omega$, as shown in Fig.~\ref{fig:map}~(c) for the case $K=10000$. The visibility reduction for linear ramps has a periodic dependence on $K$, as evident in Fig.~\ref{fig:map}~(a). The $K$-dependence occurs because atoms in different excited states of the azimuthal motion, populated by non-adiabaticity, lead to a net AI phase that varies in $K$, akin to how $\bar n$ varies with $\phi$ in the radial dynamics discussed in Sec.~\ref{sec:sims}~A and Fig.~\ref{fig:linear_sin2_compare}~(b). The visibility for linear ramps varies substantially as a function of $K$, namely between about $0.25$ and $1$,  where a value of 1 corresponds to ideal 100$\%$ visibility. In contrast, for the $\sin^2$ ramps the visibility  is practically flat near the ideal value of $1$ for all $K$. Hence, in practical implementations, one may focus only on smooth sin$^2$ lattice ramps.

Finally, we estimate the sensitivity of the interferometer. There is a trade-off between sensitivity and instrument bandwidth. The larger the number of half turns, $K$, the higher the sensitivity, but the longer it takes to acquire a sample. For a 1-Hz sampling rate or a sampling period of $\SI{1}{s}$, corresponding to a $K$-value near 2000 at a peak spin rate of $\omega_s = \SI{1}{\kHz}$, and for a phase resolution of $\delta \phi_S = 0.01$ at the inflection point of the fringes, corresponding to a number of $N=10^4$ probed atoms, Eq.~(\ref{eq:sensitivity}) yields a resolution $\delta \Omega$ of $\sim \SI{2}{nrad/s}$ for the angular platform-rotation velocity.

\section{Conclusion}

In summary, we have presented the theory and a possible design for a miniature Sagnac TAI. We propose using pairs of red-detuned, rotating azimuthal 1D optical lattices on PIC ring resonators to drag coherently-split pairs of atomic wave-packets trapped in the lattice wells. Coherently-split wave-packet components move on circular tractor trajectories with opposite rotation directions. We have assumed that axial, radial, and azimuthal dynamics are independent and that the axial motion is frozen out. The small Coriolis force that couples radial and azimuthal motion is a higher-order effect, and can be studied in future work. We have used adiabatic basis sets to reduce computational effort in the presented simulation approaches. In the adiabatic limit, which is the desired mode of TAI operation, our quantum-mechanical model agrees with the semiclassical Feynman path integral formalism. Our quantum model allows us to assess non-adiabatic effects and to thereby stake out practical operation regimes.

We have designed a PIC with COMSOL, which will be suitable to realize the presented Sagnac TAI scheme. Atom trapping in the direction transverse to the chip surface is provided by the evanescent fields of blue- and red-detuned fundamental TE00 modes in the PIC ring resonators. The trapping in the radial direction relative to the ring-resonator axes is tightened by an added blue-detuned  higher-order TE10 mode. Azimuthal trapping and dragging is accomplished by counter-propagating red-detuned TE00 modes with a well-controlled frequency difference, which form the rotating azimuthal lattices. The resultant atomic trapping potentials have been obtained. 

Based on the theory presented, we have performed quantum simulations of TAI for our exemplary PIC structure. Two types of azimuthal acceleration/deceleration schemes, namely linear and $\sin^2$ ramps, have been evaluated under conditions that approach the feasibility limits of the azimuthal-lattice rotation speed. The $\sin^2$ ramps were found to cause no discernible non-adiabatic loss of visibility of the Sagnac-TAI signal, whereas linear ramps showed a factor of up to four in visibility reduction, under otherwise identical conditions. We expect that the instrument rotation rate, $\Omega$, can be measured with a resolution of several nrad/s with a $\SI{1}{s}$ interrogation time. While this resolution is slightly below that of state-of-the-art laser or atomic gyroscopes, it would serve as proof of concept on the way towards higher resolution \cite{Dutta2016-yt}, which may be achievable with larger Sagnac areas, parallelization of multiple, identical PIC-TAIs on a single chip for improved phase resolution, and potentially with spin squeezing to evade the quantum-projection-noise limit.

When submitting our manuscript, we noted a paper that reports on a related atom-guiding methodology~\cite{Ovchinnikov}.

\section{Acknowledgments}

The work was supported by the Army Research Office and DEVCOM Army Research Laboratory under Cooperative Agreement Number W911NF-2220155. We acknowledge valuable discussions with Dr. V. Malinovsky, Dr. M. H. Goerz, Dr. S. C. Carrasco, Dr. Z. Zhang, and Dr. S. Liu.

\bibliography{bibliography.bib}

\end{document}